\documentclass[aps,prl,superscriptaddress,showpacs,twocolumn,floatfix,amsmath,amssymb]{revtex4-2}
\usepackage{hyperref} 
            
\usepackage{graphicx} 
\usepackage{xcolor}
\usepackage[version=4]{mhchem}
\usepackage{graphicx}

\def\gmn{G_{M}^n}

\def\gen{G_{E}^n}

\def\gtorder{\mathrel{\raise.3ex\hbox{$>$}\mkern-14mu
 \lower0.6ex\hbox{$\sim$}}}
\def\ltorder{\mathrel{\raise.3ex\hbox{$<$}\mkern-14mu
 \lower0.6ex\hbox{$\sim$}}}

\begin{document}
\title{A novel measurement of the neutron magnetic form factor from A=3 mirror nuclei}.

\author{S.~N.~Santiesteban}
\email{Nathaly.Santiesteban@unh.edu}
\affiliation{University of New Hampshire, Durham, New Hampshire 03824, USA} 

\author{S.~Li} \affiliation{University of New Hampshire, Durham, New Hampshire 03824, USA} \affiliation{Lawrence Berkeley National Laboratory, Berkeley, California 94720, USA}

\author{D.~Abrams} \affiliation{University of Virginia, Charlottesville, Virginia 22904, USA} 

\author{S.~Alsalmi } \affiliation{Kent State University, Kent, Ohio 44240, USA} 
\affiliation{King Saud University, Riyadh 11451, Kingdom of Saudi Arabia}

\author{D.~Androic} \affiliation{University of Zagreb, Zagreb, Croatia} 

\author{K.~Aniol} \affiliation{California State University, Los Angeles, California 90032, USA} 

\author{J.~Arrington} \affiliation{Lawrence Berkeley National Laboratory, Berkeley, California 94720, USA}
\affiliation{Argonne National Laboratory, Lemont, Illinois 60439, USA} 

\author{T.~Averett} \affiliation{William and Mary, Williamsburg, Virginia 23185, USA} 

\author{C.~Ayerbe~Gayoso} \affiliation{William and Mary, Williamsburg, Virginia 23185, USA} 

\author{J.~Bane} \affiliation{University of Tennessee, Knoxville, Tennessee 37966, USA} 

% ...

\collaboration{Jefferson Lab Hall A Collaboration}

\date{\today}

\begin{abstract}

The electromagnetic form factors of the proton and neutron encode information on the spatial structure of their charge and magnetization distributions. 
While measurements of the proton are relatively straightforward, the lack of a free neutron target makes measurements of the neutron’s electromagnetic structure more challenging and more sensitive to experimental or model-dependent uncertainties. Various experiments have attempted to extract the neutron form factors from scattering from the neutron in deuterium, with different techniques providing different, and sometimes large, systematic uncertainties. We present results from a novel measurement of the neutron magnetic form factor using quasielastic scattering from the mirror nuclei $^3$H and $^3$He, where the nuclear effects are larger than for deuterium but expected to largely cancel in the cross-section ratios. We extracted values of the neutron magnetic form factor for low-to-modest momentum transfer, $0.6<Q^2<2.9$~GeV$^2$, where existing measurements give inconsistent results. The precision and $Q^2$ range of this data allow for a better understanding of the current world's data and suggest a path toward further improvement of our overall understanding of the neutron’s magnetic form factor.
\end{abstract}
\maketitle

The proton and neutron have dual roles as both the basic building blocks of nuclei and as the lightest (nearly degenerate) baryonic bound states of QCD. Studies of their parton distribution functions provide information on the momentum distribution of the quarks inside the nucleon, while measurements of the nucleon electromagnetic form factors connect to the quarks' spatial distributions~\cite{Kelly:2002if,Miller:2007uy,Miller:2008jc}. By combining measurements on the proton and neutron, we can separate the contributions of up- and down-quarks to their internal structure.

\begin{figure}[htb]
 \includegraphics[width=0.48\textwidth, clip]{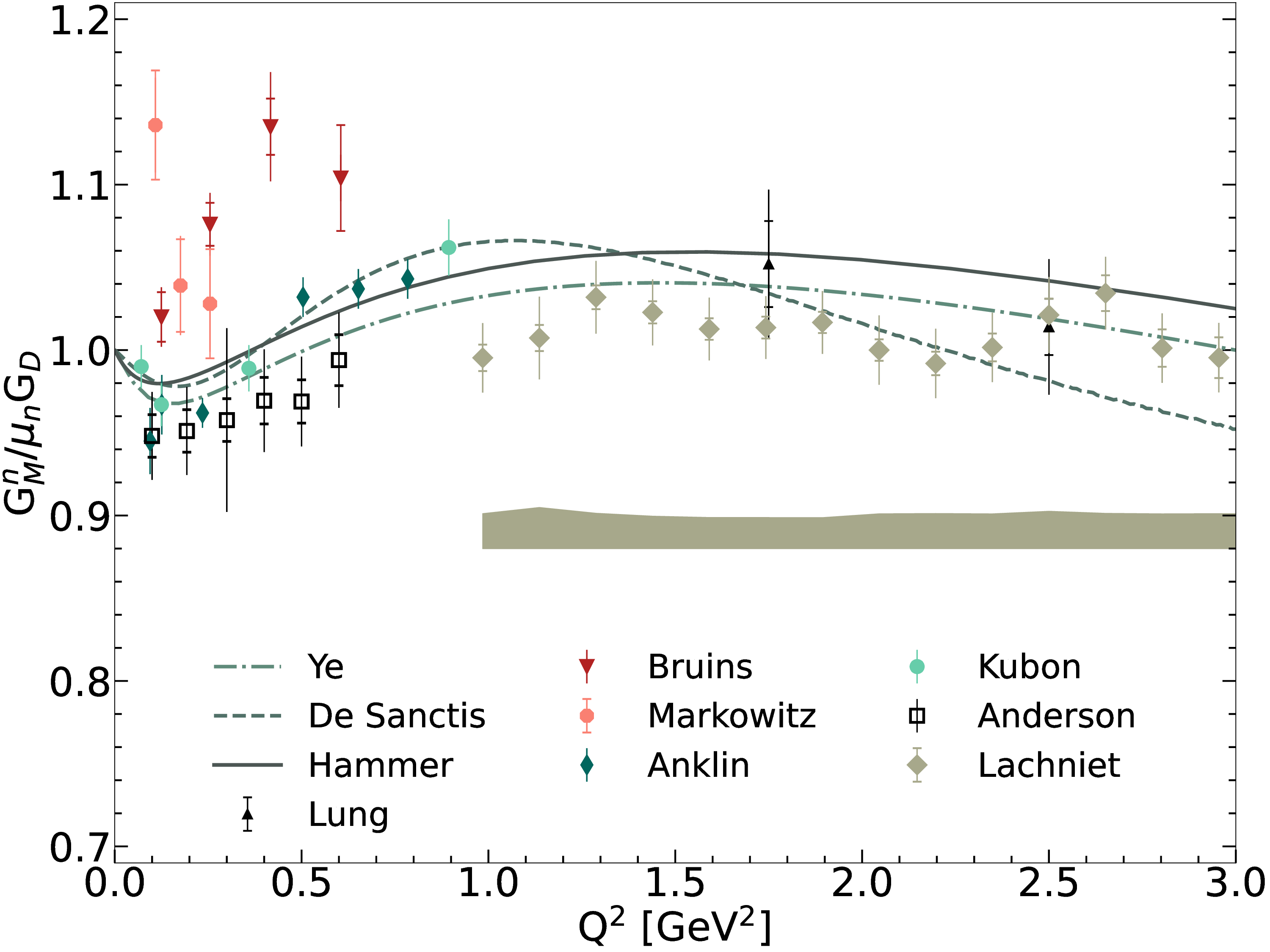}
 \caption{Previous $\gmn$ extractions~\cite{Lung:1992bu, Bruins:1995ns, Markowitz:1993hx, Anklin:1994ae, Anklin:1998ae, Kubon:2001rj, JeffersonLabE95-001:2006dax, CLAS:2008idi}, uncertainties include statistical and uncorrelated systematics, while the Lachniet result includes a band representing their correlated systematic uncertainty. The plot also shows a recent fit (Ye) of world's data~\cite{ye17}, plus curves for the hypercentral (De Sanctis) constituent quark model~\cite{DeSanctis:2005kt} and a dispersion-theoretical analysis (Hammer) from Ref.~\cite{Hammer:2003ai}.}
 \label{fig:gmnold}
\end{figure}

Because free neutrons decay with a 15-minute lifetime, measurements of neutron structure typically involve scattering from neutrons bound in nuclei, most commonly in the deuteron. For inclusive scattering, isolating the e-n elastic cross section involves correcting for the larger contribution from e-p scattering, as well as accounting for effects such as binding and Fermi motion in the nucleus~\cite{Rinat:2004xh}. Other measurements suppress the e-p contributions by measuring the neutron in the final state $^2$H(e,e'n), which requires a precise determination of the neutron detection efficiency and correcting for possible charge-exchange final-state interactions (FSI) where the struck proton scatters from the spectator neutron, which is then detected. More recently, polarized scattering from $^3$He was used to extract $\gmn$. These techniques and their limitations are discussed in Ref.~\cite{Arrington:2006zm, Perdrisat:2006hj, Arrington:2011kb}. Figure~\ref{fig:gmnold} shows several extractions of $\gmn$, divided by the neutron magnetic moment, $\mu_n$, and the dipole fit, $G_D(Q^2)=1/(1+Q^2/(0.71 GeV^2))^2$, a simple fit to the approximate $Q^2$ dependence of the nucleon form factor. While some of the more model-dependent extractions have been excluded, e.g., from low-$Q^2$ inclusive scattering, there are still large disagreements at low-$Q^2$. For some of the older measurements~\cite{Markowitz:1993hx, Bruins:1995ns}, questions have been raised about the systematic uncertainties in these extractions~\cite{Jourdan:1997pw,Bruins:1997px,Anklin:1994ae}, but even focusing on the more recent, high-precision measurements, there are discrepancies among different experiments for $0.5 < Q^2 < 1$~GeV$^2$.

We present in this work a measurement of $\gmn$ using a new technique, the comparison of quasielastic (QE) scattering from the mirror nuclei $^3$H and $^3$He, that minimizes systematic uncertainties and has small nuclear corrections. In the simplest approximation, QE scattering from a nucleus simply represents the sum of e-p and e-n elastic scattering, corrected for smearing and binding in the nucleus. Assuming nuclear corrections are similar for $^3$H and $^3$He, the scattering cross section ratio is

\begin{equation}
R = \sigma_{^3H}/\sigma_{^3He} \approx R_{Free} = (\sigma_{ep}+2\sigma_{en})/(2\sigma_{ep}+\sigma_{en}),
\end{equation}
where $R_{free}$ is the ratio neglecting nuclear effects and accounting only for the free e-N elastic cross section contributions $\sigma_{ep}$ and $\sigma_{en}$. 
Thus, the $^3$H/$^3$He cross section ratio can be expressed in terms of the $\sigma_{en}/\sigma_{ep}$ ratio, allowing for an extraction of $\sigma_{en}$, and thus $\gmn$, given our precise knowledge of $\sigma_{ep}$. However, because $\sigma_{en}/\sigma_{ep} \approx 0.4$ for these kinematics, a 1\% measurement of the $^3$H/$^3$He cross section ratio yields a 3--4\% uncertainty on $\sigma_{en}/\sigma_{ep}$. This is possible because several experimental systematic uncertainties cancel in taking the ratio $\sigma_{^3H}/\sigma_{^3He}$, and a realistic cross section model can be used to estimate the small correction difference between $R_{Free}$ and the exact $^3$H/$^3$He cross section ratio.

The experiment was performed in Hall A at Jefferson Lab (JLab) in 2018 as part of the tritium suite of experiments~\cite{Arrington:2023hht}. We used electron beam energies of 2.222 and 4.323 GeV~\cite{Santiesteban:2021lot} and detected scattered electrons in two high-resolution spectrometers (HRSs)~\cite{Alcorn:2004sb}. The basic components of the spectrometers are three superconducting quadrupoles (Q) and one superconducting dipole (D) in a QQDQ configuration. The quadrupoles focus the electrons, while the dipole disperses the electrons so their momenta can be measured.

After passing the magnets, the scattered electrons go through two Vertical Drift Chambers (VDCs) where information on the position and angle of the particles is recorded. They then pass through the trigger scintillator planes, S0 and S2, and a Cherenkov detector filled with CO$_2$ between the trigger scintillator planes. Finally, the preshower and shower lead glass blocks induce a cascade of pair production and bremsstrahlung and the energy of the particles can be measured. More detailed information about the configuration for this experiment and relevant calibrations can be found in Refs.~\cite{Li:2022fhh, cruz-torres19, cruz-torres20, santiesteban_thesis}. Because the acceptance of the spectrometer is limited to $\pm$3.5\% of the central momentum, multiple HRS momentum settings were used to more completely cover the QE peak. The kinematics, number of settings, and extracted form factors are shown in Table~\ref{tab:kin}.

The $^3$H, $^3$He, and $^2$H gas targets were contained in 25~cm long aluminum cells~\cite{target_NIM, Arrington:2023hht}. Two $^3$H cells from Savannah River Site were used, one for each run period~\cite{Li:2022fhh}. The $^3$He target thicknes is 53.23$\pm$0.53~mg/cm$^2$, while the two $^3$H cells had nominal thicknesses of 84.95$\pm$0.28 and 84.79$\pm$0.28~mg/cm$^2$, before accounting for $^3$H decay or target density modifications due to beam heating. It was found that the second cell, used for the data taking at 4.323~GeV, had a (4.12$\pm$0.20)\% $^1$H contamination~\cite{Li:2022fhh}. To correct for this contamination, all settings with $^1$H elastic data ware used to estimate the amount of contamination an simulations were used to subtract the $^1$H contribution from the $^3$H data. The $^3$H thickness was then reduced by (4.12$\pm$0.20)\% to correct for the presence of $^1$H~\cite{Li:2022fhh}. A correction was also applied to account for the reduction in gas density seen by the beam as a result of target heating, determined to be a 9.4\% (6.0\%) for $^3$H ($^3$He)~\cite{Santiesteban:2018qwi} at the average beam current of the experiment. Finally, because the tritium decays into $^3$He over time (up to 4.21\% by the end of the run period), the $^3$He contribution was subtracted, based on the $^3$He measurement, and the target thickness was reduced to account for the tritium decay.  

\begin{table}
\begin{tabular}{c c c c c c}
\hline \hline
Label   & $E_0$     & ~Theta~    & $Q^2$   &  \# QE  & $\gmn / (\mu_n G_D$) \\ 
        & [GeV]     & [deg]    & [GeV$^2$]     & settings  &        \\\hline
L21     &2.222      & 21.778        &  0.603        &   3       & 1.066$\pm$0.017$\pm$0.027~\\
L24     &2.222      & 23.891        &  0.703        &   3       & 1.049$\pm$0.016$\pm$0.026\\
L26     &2.222      & 25.952        &  0.803        &   3       & 1.067$\pm$0.017$\pm$0.026\\
L28     &2.222      & 28.001        &  0.905        &   3       & 1.052$\pm$0.017$\pm$0.025\\  
L30     &2.222      & 30.001        &  1.004        &   3       & 1.058$\pm$0.017$\pm$0.025\\
L17     &4.323      & 17.006        &  1.360        &   2       & 1.039$\pm$0.018$\pm$0.025\\
R42     &2.222      & 42.025        &  1.578        &   3       & 1.067$\pm$0.028$\pm$0.024\\
R24     &4.323      & 24.016        &  2.313        &   2       & 1.068$\pm$0.022$\pm$0.025\\
R26     &4.323      & 26.003        &  2.580        &   3       & 1.034$\pm$0.023$\pm$0.025\\
R28     &4.323      & 28.004        &  2.843        &   2       & 1.021$\pm$0.029$\pm$0.025\\
\hline \hline
\end{tabular}
\caption{Kinematics including the number of settings used to cover the QE peak, and the extracted form factor and uncertainties (uncorrelated and correlated).}
\label{tab:kin}
\end{table}

Cuts were applied to the reconstructed angle and momentum of the scattered electrons to focus on the high-acceptance regions of spectrometers. The small pion contribution was removed by applying cuts to the Cherenkov and shower counter detectors, yielding a negligible ($<$0.1\%) pion contribution~\cite{santiesteban_thesis}. To subtract the large contribution from the target endcaps, the reaction vertex was selected to be $\pm$8~cm from the center of the target and the small ($<$1\%) residual contribution from the endcaps was removed using data from an empty cell or dummy target (two thicker Al foils at the position of the target windows that were used when the rate was low), as described in Ref.~\cite{Li:2022fhh}.

After applying cuts, the yield was normalized to the effective integrated luminosity, which includes the target length, the data acquisition live time, the trigger, tracking, and particle identification efficiencies. The normalized yield was binned as a function of energy transfer, $\omega$, and compared to a detailed simulation of the experiment. The simulation generates events over the acceptance of the spectrometer, weighted with a realistic cross section model that starts with a model of the Born cross section and then accounts for energy loss, multiple scattering, and radiative corrections~\cite{Dasu:1993vk}, as described in~\cite{Arrington:2021vuu}. The events were then propagated through a model of the HRS spectrometer to account for the spectrometer acceptance.

For each bin in $\omega$, we took the cross section model and scaled it by the ratio of the normalized yield from the experiment to the normalized yield from the simulation. Assuming that the simulation accounts for all of the corrections needed to go from the Born cross section to the observed number of events, the only remaining uncertainty in the simulation was the model cross section itself, and this procedure adjusts the model on a bin-by-bin basis to reproduce the data. In this procedure, any imperfections in the simulation (radiative corrections, acceptance, etc.) could modify the cross section, and we evaluated each of the aspects of the simulation to account for these uncertainties~\cite{santiesteban_thesis}. As discussed in the following sections, the main observable we are interested in for the extraction of $\gmn$ is the ratio of $^3$H and $^3$He cross sections, integrated over the QE peak. The extracted cross sections have an estimated point-to-point systematic uncertainty of 1.8--2.8\% and a normalization uncertainty of roughly 3\%. In taking this ratio, many sources of uncertainty, including most of the largest ones, cancel out and we are left with a much smaller systematic uncertainty. We note that for the R42 setting, $Q^2\approx 1.6$~GeV$^2$, the cross sections for $^3$H, $^3$He, and $^2$H were all about 15\% below our simple QE cross section model. Because this was the largest angle, the spectrometer saw the largest effective target length, and the fact that target length acceptance wasn't sufficiently well reconstructed for long targets in the Monte Carlo led to a reduced cross section. This effect should cancel out in the ratios, and we tested this by comparing the ratio with the standard cut and with a $\pm$4~cm cut. The tighter cut raised the absolute cross sections for all targets but had minimal impact on the various cross section ratios (typically 0.5\%). For the extraction of the QE cross section ratio, we treat this data set consistently with all the others and apply an additional 1\% uncertainty in the ratio to account for the possible target-dependent impact of the imperfect modeling of the target length acceptance.

\begin{figure}[htb]
 \includegraphics[width=0.48\textwidth, height=5.5cm, trim={0mm 0mm -5mm 0mm}, clip]{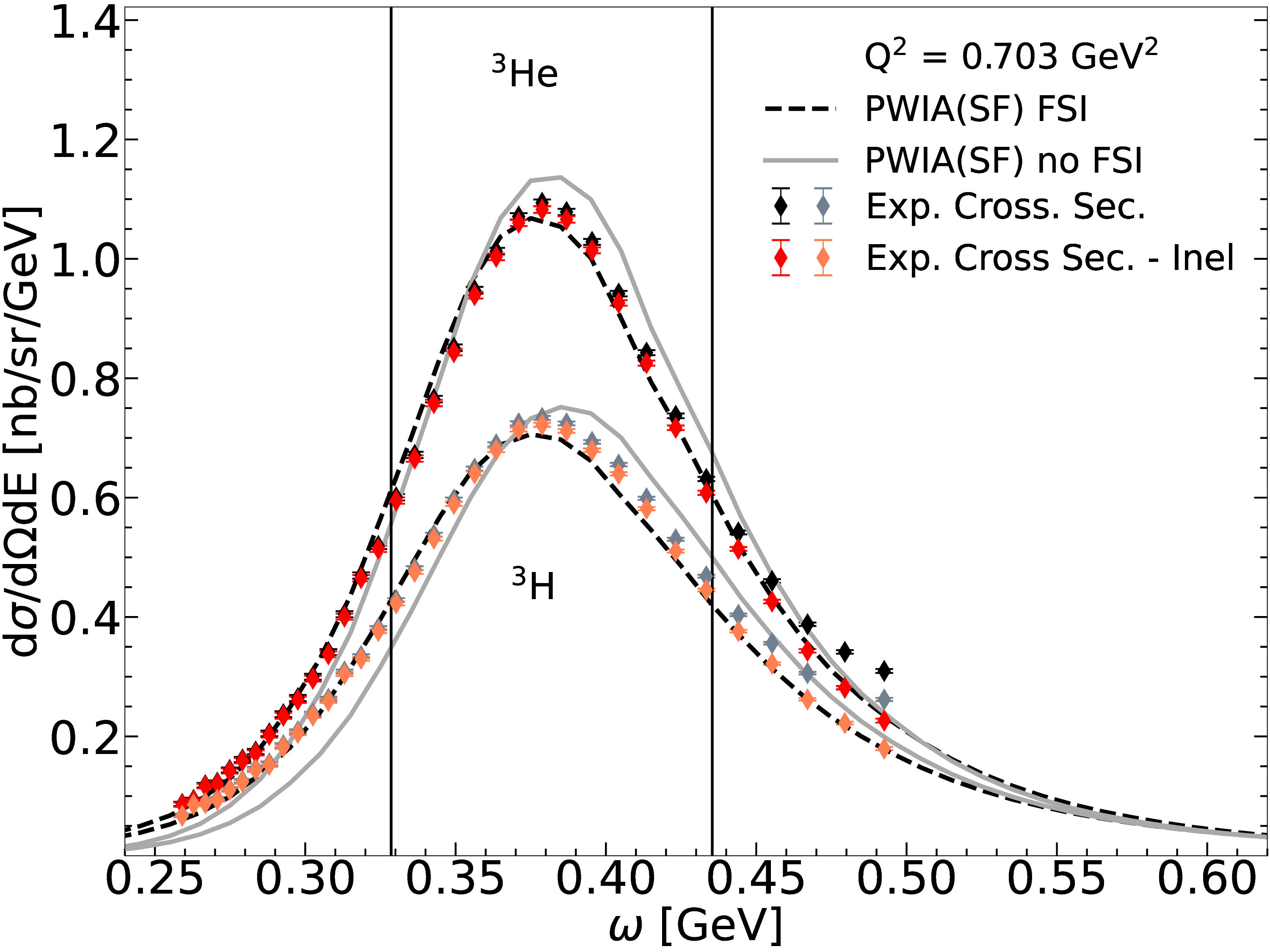}
 \includegraphics[width=0.48\textwidth, height=5.5cm, trim={0mm 0mm -5mm 0mm}, clip]{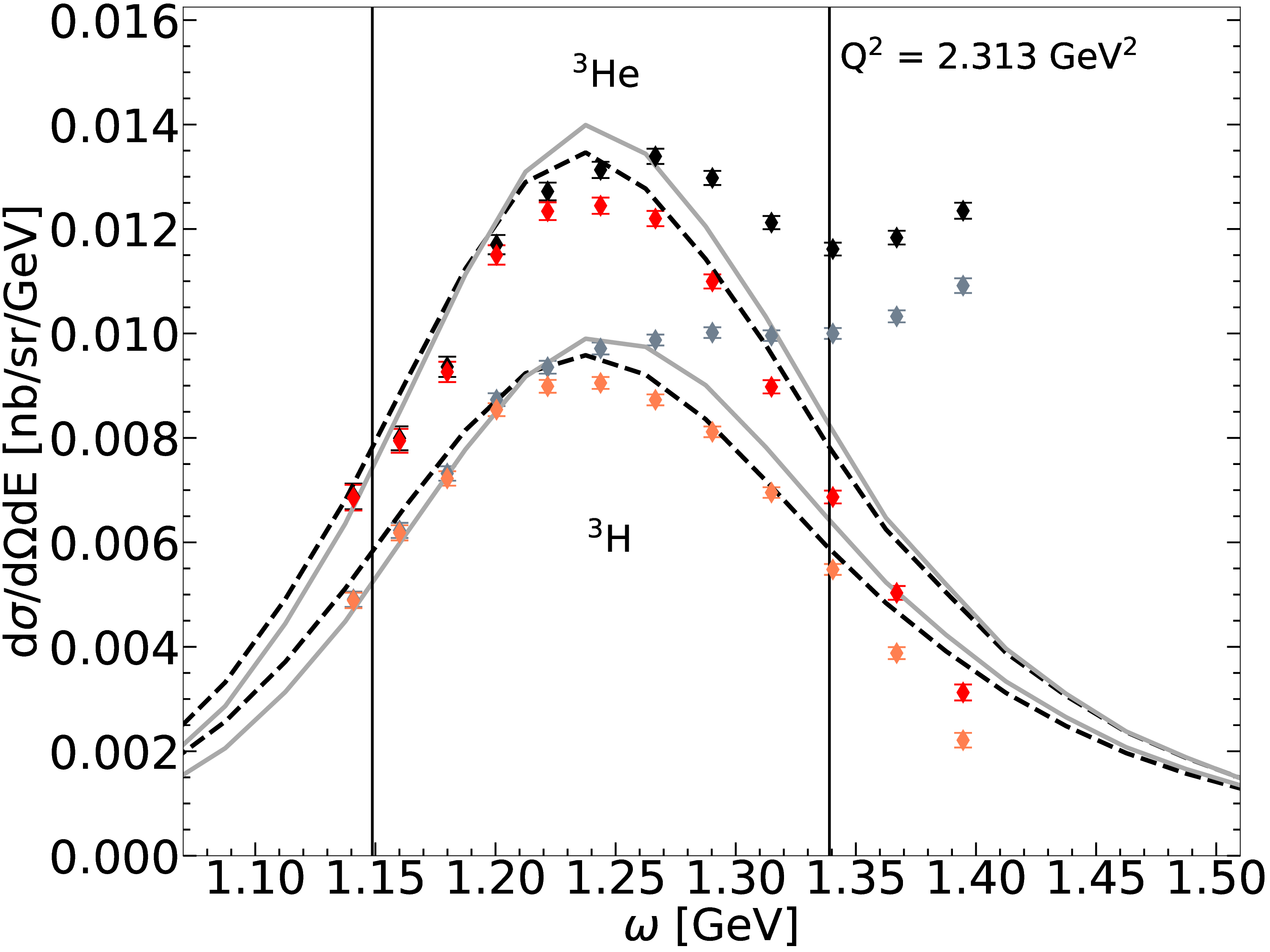}
 \caption{Cross sections and statistical uncertainties for $^3$H and $^3$He compared to short-time approximation calculations of Ref.~\cite{Andreoli:2021cxo} for the L24 (0.703 GeV$^2$) and R24 (2.313 GeV$^2$) settings. The black (grey) points are the measured total cross section, and the red (orange) points are after subtraction of the inelastic contribution. The vertical lines represent $\pm$1$\sigma$ from the QE peak (see text for details).}
 \label{fig:calcs}
\end{figure}

Figure~\ref{fig:calcs} shows the $^3$H and $^3$He cross sections from settings L24 and R24 with calculations based on Ref~\cite{Andreoli:2021cxo}. The calculations that include FSI were used in the $\gmn$ extraction. On the high $\omega$ side, we have subtracted the inelastic contribution using the model of Ref.~\cite{Bosted:2012qc} but with a modified meson-exchange contribution (MEC) (discussed below). Even where the subtraction is large, the inelastic-subtracted result is in fairly good agreement with the calculation, and because the subtraction is similar for both targets, the impact of the inelastic subtraction on the cross section is smaller when taking the ratio.

For the $\gmn$ extraction, we integrated over the QE peak using only the statistical uncertainties, take the $^3$H/$^3$He ratio, and then apply the contribution of uncertainties that do not cancel. The point-to-point systematic uncertainty is $\sim$0.75\%, where the major sources are model and cut dependence in the cross section extraction, radiative corrections, inelastic and MEC subtraction, and $^3$He contamination. The normalization uncertainty, common for all $Q^2$ values, is 1.2\%, dominated by the 1.08\% target thickness uncertainty. While two different $^3$H targets were used, the uncertainty is dominated by our knowledge of the equation of state and calibration of the pressure and temperature measurements, which were identical for both run periods.

We integrate over the central region of the peak for both nuclei, and take the $^3$H/$^3$He cross section ratio. We choose an integration range of $\pm$1 standard deviation (as determined by a gaussian fit to the calculations) to minimize our sensitivity to any disagreement in the low-$\omega$ tail and to the inelastic subtraction which is larger on the high-$\omega$ side. We also apply a small offset in $\omega$ to the calculations for both targets, so that the peak positions are consistent with the data. This way we ensure that we are integrating around the center of the QE peak for both data and simulation. This, combined with the symmetric integration region around the peak, minimizes sensitivity to any residual offset. 

As noted earlier, the cross section ratio is approximately $R_{Free} = (1+2\sigma_{en}/\sigma_{ep}) / (2+\sigma_{en}/\sigma_{ep})$, allowing for an extraction of $\sigma_{en}/\sigma_{ep}$. There is a small correction factor, $\alpha$, that accounts for the difference in nuclear effects and the impact of integrating the QE peak over a finite range ($R=\alpha R_{Free}$). We use cross section calculations~\cite{Andreoli:2021cxo} to determine $\alpha$, the difference between this approximation and the full QE cross section ratio, integrated over the central part of the QE peak. The impact of off-shell effects is also accounted for in the extraction of $\alpha$ from these calculations, but they are a very small correction as the n/p cross section ratio is much less modified by off-shell effects than the individual cross sections.

To estimate the impact of changing the integration region, we expand the high- or low-$\omega$ from 1 to 1.5~$\sigma$. In cases where there is insufficient data to expand the cut, or where this includes data where the inelastic subtraction is very large, we use a tighter cut instead to estimate the dependence. We observe a typical variation of 0.3\% in $\sigma_{en}/\sigma_{ep}$, corresponding to less than a 0.1\% change to the the $^3$H/$^3$He ratio. We apply an additional 0.1\% point-to-point and 0.1\% normalization uncertainty to the $^3$H/$^3$He ratio to account for the cut dependence.  

The functional form of the MEC to the inelastic model~\cite{Bosted:2012qc} was not intended to cover the low-$\omega$ side of the QE peak, and gives an unrealistically large contribution, especially at low $Q^2$. To avoid a large over-subtraction, we modified MEC contribution using different cutoff functions that reduced the low-$\omega$ contributions, as described in the Supplemental Material~\cite{supplemental}. We compared these results to subtractions using no MEC and calculations~\cite{priv_MEC} based on Ref.~\cite{PhysRevC.104.025501}. While the calculated MEC were smaller than our modified parmaterization, they yielded a somewhat larger correction due to the difference in the isospin structure. For the final results, we take the ratio based on our intermediate truncated MEC parameterization~\cite{supplemental}, applying a 100\% uncertainty on the MEC subtraction, which roughly covers the range of all of the methods discussed above.

We extract $\sigma_{en}$ by multiplying the extracted value of $\sigma_{en}/\sigma_{ep}$ by the proton cross section from the parameterization of Ref.~\cite{arrington:2007ux} that does not include corrections for two-photon exchange (TPE), taking a 1\% uncertainty on the value of $\sigma_{ep}$. We then apply TPE corrections to the extracted $\sigma_{en}$, based on the calculations from Ref.~\cite{Arrington:2011dn} (0.5\% for these kinematics) to obtain $\sigma_{en}$ in the Born approximation. We subtract the contribution to the elastic cross section from $\gen$ (typically 5\% of the total) using the value and uncertainty of $\gen$ from Ref.~\cite{ye17} to obtain $\gmn$.

\begin{figure}[htb]
 \includegraphics[width=0.48\textwidth, clip]{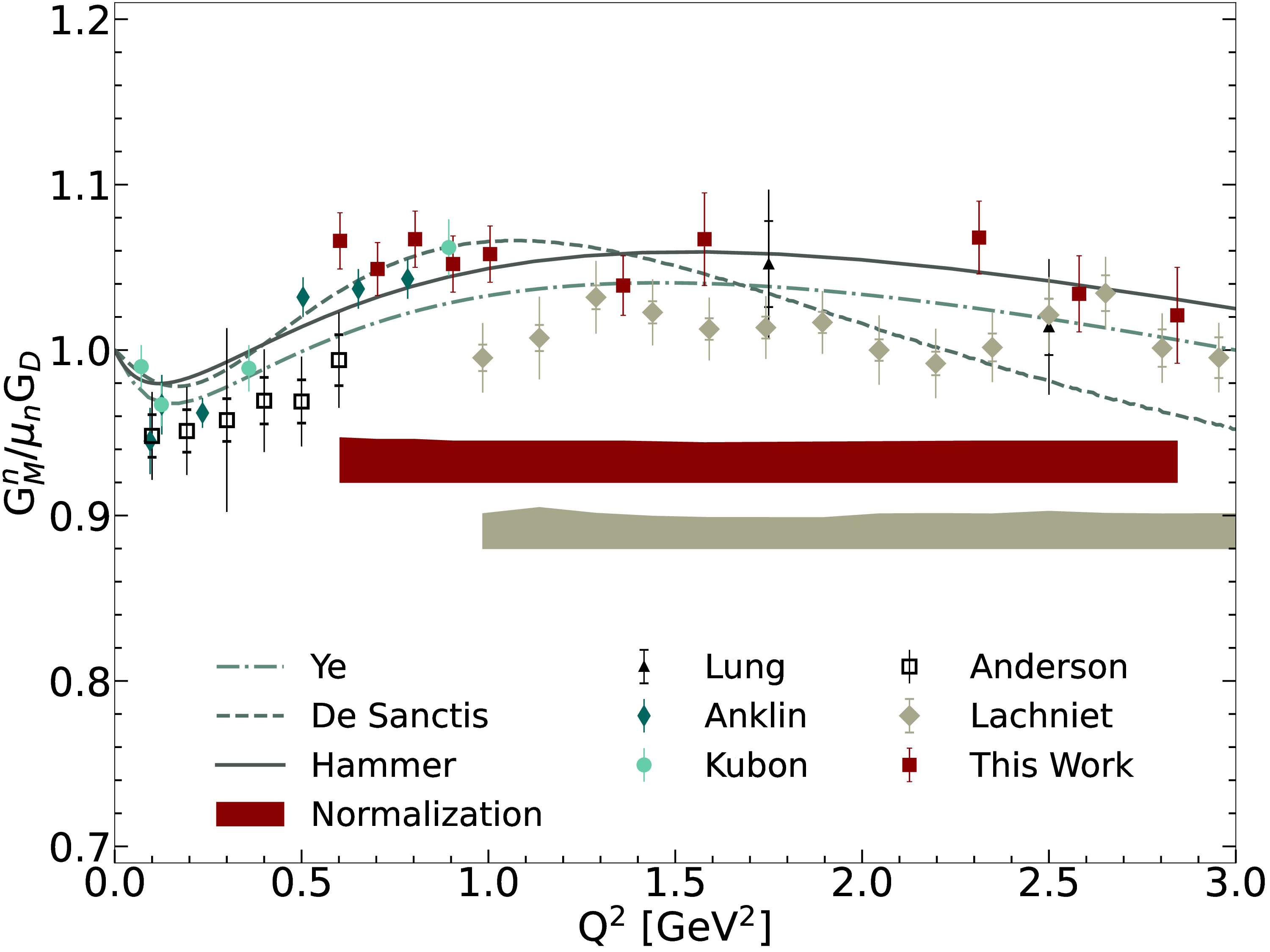}
 \caption{Our new $\gmn$ results along with a subset of previous measurements~\cite{Kubon:2001rj,Anklin:1998ae,JeffersonLabE95-001:2006dax,CLAS:2008idi} (see text for details).}
 \label{fig:gmnnew}
\end{figure}

The extracted values of $\gmn$ are shown in Fig.~\ref{fig:gmnnew}, along with a subset of previous measurements including the highest precision data sets and measurements covering a significant $Q^2$ range. These are the data sets we use in the fit described below, and data sets with only one or two points and large uncertainties do not contribute significantly to the fit. Our results are in good agreement with the Mainz extractions~\cite{Anklin:1998ae, Kubon:2001rj}, and somewhat higher than previous JLab extractions from polarization~\cite{JeffersonLabE95-001:2006dax} and cross section ratio~\cite{CLAS:2008idi} extractions. However, given our correlated uncertainty of approximately 2.4\%, our results are only 2$\sigma$ above these experiments.  

An important question is whether or not these data sets yield consistent results when taking into account all of the uncertainties, in particular highly-correlated uncertainties that would allow full data sets to shift up or down. We note that the previous extractions shown in Fig.~\ref{fig:gmnnew} typically include only statistical and uncorrelated systematic uncertainties, though it is likely that several of these uncertainties have a strongly-correlated component. These could include corrections associated with the proton and neutron detection efficiencies, detector acceptance, and radiative corrections. Beyond experimental corrections, extracting $\gmn$ from the cross section ratios or polarization observables requires models for nuclear effects, knowledge of the electron-proton cross section, final-state interactions, off-shell effects, and hard TPE effects~\cite{Arrington:2011dn}. Based on an examination of the dominant uncertainties in these works, we estimate that these experiments have correlated uncertainties on their extracted values of $\gmn$ that vary between 1.4-2.0\%.

To examine the significance associated with these correlated uncertainties, we perform a global fit after applying a 1.5\% normalization uncertainty on each of the previous data sets, based on a rough estimate of their correlated experimental and model-dependent uncertainties. We fit $\gmn$ to a 4rd order inverse polynomial, neglecting the normalization uncertainties, and obtain $\chi^2=74.7$ for 46 degrees of freedom. If we allow the normalization of each data set to vary, with a $\chi^2$ penalty based on its normalization uncertainty, the fit gives $\chi^2=47.0$ for 46 degrees of freedom, bringing the data sets into good agreement with normalizations shifts that are below 1$\sigma$, except for the CLAS data set which is raised by 1.3$\sigma$.  

When we account for the estimated normalization uncertainties for the various data sets, we find that they are in excellent agreement. While our normalization uncertainty is somewhat larger than we estimated for the previous measurements, the fact that our data set has overlap with the Anklin, Kubon, Lachniet, and Anderson extractions allow it to provide an improved cross-normalization of the various data sets. 

In conclusion, we have extracted $\gmn$ for $Q^2$ values from 0.6-2.9~GeV$^2$, with point-to-point uncertainties of 1.5--2\% and an additional correlated uncertainty of approximately 2.4\%. Part of the normalization uncertainty comes from our subtraction of the MEC, so this extraction can be improved with better understanding of the MECs. The current uncertainties, combined with the $Q^2$ coverage of these data, allow us to better constrain the normalization of various data sets. This led us to reexamine the correlated uncertainties in previous data sets, and demonstrates that the data sets are consistent within their uncertainties, taking our estimate of 1.5\% normalization uncertainty for the previous measurements. This suggests that overall understanding of $\gmn$ could be further improved with the inclusion of data sets covering a large $Q^2$ range, even with a significant normalization uncertainty, or with the addition of a new highly precise and accurate measurement, even with a very limited $Q^2$ range or single $Q^2$ point. 

\begin{acknowledgments}

We acknowledge useful discussions with J. Amaro, O. Benhar, and M. Sargsian and the contribution of the JLab target group and technical staff for design and construction of the Tritium target and their support running this experiment. This work was supported in part by the Department of Energy's Office of Science, Office of Nuclear Physics, under contracts DE-AC02-05CH11231, DE-FG02-88ER40410, DE-SC0014168, DE-FG02-96ER40950, and DE-SC0014168, the National Science Foundation, including grant NSF PHY-1714809, the Science Committee of Republic of Armenia under grant 21AG-1C085, and the Tsinghua University Initiative Scientific Research Program, and DOE contract DE-AC05-06OR23177 under which JSA, LLC operates JLab. 

\end{acknowledgments}

\bibliographystyle{apsrev4-1}

\bibliography{tritium-QE}

\begin{thebibliography}{38}%
\makeatletter
\providecommand \@ifxundefined [1]{%
 \@ifx{#1\undefined}
}%
\providecommand \@ifnum [1]{%
 \ifnum #1\expandafter \@firstoftwo
 \else \expandafter \@secondoftwo
 \fi
}%
\providecommand \@ifx [1]{%
 \ifx #1\expandafter \@firstoftwo
 \else \expandafter \@secondoftwo
 \fi
}%
\providecommand \natexlab [1]{#1}%
\providecommand \enquote  [1]{``#1''}%
\providecommand \bibnamefont  [1]{#1}%
\providecommand \bibfnamefont [1]{#1}%
\providecommand \citenamefont [1]{#1}%
\providecommand \href@noop [0]{\@secondoftwo}%
\providecommand \href [0]{\begingroup \@sanitize@url \@href}%
\providecommand \@href[1]{\@@startlink{#1}\@@href}%
\providecommand \@@href[1]{\endgroup#1\@@endlink}%
\providecommand \@sanitize@url [0]{\catcode `\\12\catcode `\$12\catcode
  `\&12\catcode `\#12\catcode `\^12\catcode `\_12\catcode `\%12\relax}%
\providecommand \@@startlink[1]{}%
\providecommand \@@endlink[0]{}%
\providecommand \url  [0]{\begingroup\@sanitize@url \@url }%
\providecommand \@url [1]{\endgroup\@href {#1}{\urlprefix }}%
\providecommand \urlprefix  [0]{URL }%
\providecommand \Eprint [0]{\href }%
\providecommand \doibase [0]{http://dx.doi.org/}%
\providecommand \selectlanguage [0]{\@gobble}%
\providecommand \bibinfo  [0]{\@secondoftwo}%
\providecommand \bibfield  [0]{\@secondoftwo}%
\providecommand \translation [1]{[#1]}%
\providecommand \BibitemOpen [0]{}%
\providecommand \bibitemStop [0]{}%
\providecommand \bibitemNoStop [0]{.\EOS\space}%
\providecommand \EOS [0]{\spacefactor3000\relax}%
\providecommand \BibitemShut  [1]{\csname bibitem#1\endcsname}%
\let\auto@bib@innerbib\@empty
\bibitem [{\citenamefont {Kelly}(2002)}]{Kelly:2002if}%
  \BibitemOpen
  \bibfield  {author} {\bibinfo {author} {\bibfnamefont {J.~J.}\ \bibnamefont
  {Kelly}},\ }\href {\doibase 10.1103/PhysRevC.66.065203} {\bibfield  {journal}
  {\bibinfo  {journal} {Phys. Rev. C}\ }\textbf {\bibinfo {volume} {66}},\
  \bibinfo {pages} {065203} (\bibinfo {year} {2002})},\ \Eprint
  {http://arxiv.org/abs/hep-ph/0204239} {arXiv:hep-ph/0204239} \BibitemShut
  {NoStop}%
\bibitem [{\citenamefont {Miller}(2007)}]{Miller:2007uy}%
  \BibitemOpen
  \bibfield  {author} {\bibinfo {author} {\bibfnamefont {G.~A.}\ \bibnamefont
  {Miller}},\ }\href {\doibase 10.1103/PhysRevLett.99.112001} {\bibfield
  {journal} {\bibinfo  {journal} {Phys. Rev. Lett.}\ }\textbf {\bibinfo
  {volume} {99}},\ \bibinfo {pages} {112001} (\bibinfo {year} {2007})},\
  \Eprint {http://arxiv.org/abs/0705.2409} {arXiv:0705.2409 [nucl-th]}
  \BibitemShut {NoStop}%
\bibitem [{\citenamefont {Miller}\ and\ \citenamefont
  {Arrington}(2008)}]{Miller:2008jc}%
  \BibitemOpen
  \bibfield  {author} {\bibinfo {author} {\bibfnamefont {G.~A.}\ \bibnamefont
  {Miller}}\ and\ \bibinfo {author} {\bibfnamefont {J.}~\bibnamefont
  {Arrington}},\ }\href {\doibase 10.1103/PhysRevC.78.032201} {\bibfield
  {journal} {\bibinfo  {journal} {Phys. Rev. C}\ }\textbf {\bibinfo {volume}
  {78}},\ \bibinfo {pages} {032201} (\bibinfo {year} {2008})},\ \Eprint
  {http://arxiv.org/abs/0806.3977} {arXiv:0806.3977 [nucl-th]} \BibitemShut
  {NoStop}%
\bibitem [{\citenamefont {Lung}\ \emph {et~al.}(1993)\citenamefont {Lung} \emph
  {et~al.}}]{Lung:1992bu}%
  \BibitemOpen
  \bibfield  {author} {\bibinfo {author} {\bibfnamefont {A.}~\bibnamefont
  {Lung}} \emph {et~al.},\ }\href {\doibase 10.1103/PhysRevLett.70.718}
  {\bibfield  {journal} {\bibinfo  {journal} {Phys. Rev. Lett.}\ }\textbf
  {\bibinfo {volume} {70}},\ \bibinfo {pages} {718} (\bibinfo {year}
  {1993})}\BibitemShut {NoStop}%
\bibitem [{\citenamefont {Bruins}\ \emph {et~al.}(1995)\citenamefont {Bruins}
  \emph {et~al.}}]{Bruins:1995ns}%
  \BibitemOpen
  \bibfield  {author} {\bibinfo {author} {\bibfnamefont {E.~E.}\ \bibnamefont
  {Bruins}} \emph {et~al.},\ }\href {\doibase 10.1103/PhysRevLett.75.21}
  {\bibfield  {journal} {\bibinfo  {journal} {Phys. Rev. Lett.}\ }\textbf
  {\bibinfo {volume} {75}},\ \bibinfo {pages} {21} (\bibinfo {year}
  {1995})}\BibitemShut {NoStop}%
\bibitem [{\citenamefont {Markowitz}\ \emph {et~al.}(1993)\citenamefont
  {Markowitz} \emph {et~al.}}]{Markowitz:1993hx}%
  \BibitemOpen
  \bibfield  {author} {\bibinfo {author} {\bibfnamefont {P.}~\bibnamefont
  {Markowitz}} \emph {et~al.},\ }\href {\doibase 10.1103/PhysRevC.48.R5}
  {\bibfield  {journal} {\bibinfo  {journal} {Phys. Rev. C}\ }\textbf {\bibinfo
  {volume} {48}},\ \bibinfo {pages} {R5} (\bibinfo {year} {1993})}\BibitemShut
  {NoStop}%
\bibitem [{\citenamefont {Anklin}\ \emph {et~al.}(1994)\citenamefont {Anklin}
  \emph {et~al.}}]{Anklin:1994ae}%
  \BibitemOpen
  \bibfield  {author} {\bibinfo {author} {\bibfnamefont {H.}~\bibnamefont
  {Anklin}} \emph {et~al.},\ }\href {\doibase 10.1016/0370-2693(94)90538-X}
  {\bibfield  {journal} {\bibinfo  {journal} {Phys. Lett. B}\ }\textbf
  {\bibinfo {volume} {336}},\ \bibinfo {pages} {313} (\bibinfo {year}
  {1994})}\BibitemShut {NoStop}%
\bibitem [{\citenamefont {Anklin}\ \emph {et~al.}(1998)\citenamefont {Anklin}
  \emph {et~al.}}]{Anklin:1998ae}%
  \BibitemOpen
  \bibfield  {author} {\bibinfo {author} {\bibfnamefont {H.}~\bibnamefont
  {Anklin}} \emph {et~al.},\ }\href {\doibase 10.1016/S0370-2693(98)00442-0}
  {\bibfield  {journal} {\bibinfo  {journal} {Phys. Lett. B}\ }\textbf
  {\bibinfo {volume} {428}},\ \bibinfo {pages} {248} (\bibinfo {year}
  {1998})}\BibitemShut {NoStop}%
\bibitem [{\citenamefont {Kubon}\ \emph {et~al.}(2002)\citenamefont {Kubon}
  \emph {et~al.}}]{Kubon:2001rj}%
  \BibitemOpen
  \bibfield  {author} {\bibinfo {author} {\bibfnamefont {G.}~\bibnamefont
  {Kubon}} \emph {et~al.},\ }\href {\doibase 10.1016/S0370-2693(01)01386-7}
  {\bibfield  {journal} {\bibinfo  {journal} {Phys. Lett. B}\ }\textbf
  {\bibinfo {volume} {524}},\ \bibinfo {pages} {26} (\bibinfo {year} {2002})},\
  \Eprint {http://arxiv.org/abs/nucl-ex/0107016} {arXiv:nucl-ex/0107016}
  \BibitemShut {NoStop}%
\bibitem [{\citenamefont {Anderson}\ \emph {et~al.}(2007)\citenamefont
  {Anderson} \emph {et~al.}}]{JeffersonLabE95-001:2006dax}%
  \BibitemOpen
  \bibfield  {author} {\bibinfo {author} {\bibfnamefont {B.}~\bibnamefont
  {Anderson}} \emph {et~al.} (\bibinfo {collaboration} {Jefferson Lab
  E95-001}),\ }\href {\doibase 10.1103/PhysRevC.75.034003} {\bibfield
  {journal} {\bibinfo  {journal} {Phys. Rev. C}\ }\textbf {\bibinfo {volume}
  {75}},\ \bibinfo {pages} {034003} (\bibinfo {year} {2007})},\ \Eprint
  {http://arxiv.org/abs/nucl-ex/0605006} {arXiv:nucl-ex/0605006} \BibitemShut
  {NoStop}%
\bibitem [{\citenamefont {Lachniet}\ \emph {et~al.}(2009)\citenamefont
  {Lachniet} \emph {et~al.}}]{CLAS:2008idi}%
  \BibitemOpen
  \bibfield  {author} {\bibinfo {author} {\bibfnamefont {J.}~\bibnamefont
  {Lachniet}} \emph {et~al.} (\bibinfo {collaboration} {CLAS}),\ }\href
  {\doibase 10.1103/PhysRevLett.102.192001} {\bibfield  {journal} {\bibinfo
  {journal} {Phys. Rev. Lett.}\ }\textbf {\bibinfo {volume} {102}},\ \bibinfo
  {pages} {192001} (\bibinfo {year} {2009})}\BibitemShut {NoStop}%
\bibitem [{\citenamefont {Ye}\ \emph {et~al.}(2018)\citenamefont {Ye},
  \citenamefont {Arrington}, \citenamefont {Hill},\ and\ \citenamefont
  {Lee}}]{ye17}%
  \BibitemOpen
  \bibfield  {author} {\bibinfo {author} {\bibfnamefont {Z.}~\bibnamefont
  {Ye}}, \bibinfo {author} {\bibfnamefont {J.}~\bibnamefont {Arrington}},
  \bibinfo {author} {\bibfnamefont {R.~J.}\ \bibnamefont {Hill}}, \ and\
  \bibinfo {author} {\bibfnamefont {G.}~\bibnamefont {Lee}},\ }\href {\doibase
  https://doi.org/10.1016/j.physletb.2017.11.023} {\bibfield  {journal}
  {\bibinfo  {journal} {Physics Letters B}\ }\textbf {\bibinfo {volume}
  {777}},\ \bibinfo {pages} {8} (\bibinfo {year} {2018})}\BibitemShut {NoStop}%
\bibitem [{\citenamefont {De~Sanctis}\ \emph {et~al.}(2007)\citenamefont
  {De~Sanctis}, \citenamefont {Giannini}, \citenamefont {Santopinto},\ and\
  \citenamefont {Vassallo}}]{DeSanctis:2005kt}%
  \BibitemOpen
  \bibfield  {author} {\bibinfo {author} {\bibfnamefont {M.}~\bibnamefont
  {De~Sanctis}}, \bibinfo {author} {\bibfnamefont {M.~M.}\ \bibnamefont
  {Giannini}}, \bibinfo {author} {\bibfnamefont {E.}~\bibnamefont
  {Santopinto}}, \ and\ \bibinfo {author} {\bibfnamefont {A.}~\bibnamefont
  {Vassallo}},\ }\href {\doibase 10.1103/PhysRevC.76.062201} {\bibfield
  {journal} {\bibinfo  {journal} {Phys. Rev. C}\ }\textbf {\bibinfo {volume}
  {76}},\ \bibinfo {pages} {062201} (\bibinfo {year} {2007})},\ \Eprint
  {http://arxiv.org/abs/nucl-th/0506033} {arXiv:nucl-th/0506033} \BibitemShut
  {NoStop}%
\bibitem [{\citenamefont {Hammer}\ and\ \citenamefont
  {Meissner}(2004)}]{Hammer:2003ai}%
  \BibitemOpen
  \bibfield  {author} {\bibinfo {author} {\bibfnamefont {H.~W.}\ \bibnamefont
  {Hammer}}\ and\ \bibinfo {author} {\bibfnamefont {U.-G.}\ \bibnamefont
  {Meissner}},\ }\href {\doibase 10.1140/epja/i2003-10223-y} {\bibfield
  {journal} {\bibinfo  {journal} {Eur. Phys. J. A}\ }\textbf {\bibinfo {volume}
  {20}},\ \bibinfo {pages} {469} (\bibinfo {year} {2004})},\ \Eprint
  {http://arxiv.org/abs/hep-ph/0312081} {arXiv:hep-ph/0312081} \BibitemShut
  {NoStop}%
\bibitem [{\citenamefont {Rinat}\ \emph {et~al.}(2007)\citenamefont {Rinat},
  \citenamefont {Taragin},\ and\ \citenamefont {Viviani}}]{Rinat:2004xh}%
  \BibitemOpen
  \bibfield  {author} {\bibinfo {author} {\bibfnamefont {A.~S.}\ \bibnamefont
  {Rinat}}, \bibinfo {author} {\bibfnamefont {M.~F.}\ \bibnamefont {Taragin}},
  \ and\ \bibinfo {author} {\bibfnamefont {M.}~\bibnamefont {Viviani}},\ }\href
  {\doibase 10.1016/j.nuclphysa.2006.10.086} {\bibfield  {journal} {\bibinfo
  {journal} {Nucl. Phys. A}\ }\textbf {\bibinfo {volume} {784}},\ \bibinfo
  {pages} {25} (\bibinfo {year} {2007})}\BibitemShut {NoStop}%
\bibitem [{\citenamefont {Arrington}\ \emph
  {et~al.}(2007{\natexlab{a}})\citenamefont {Arrington}, \citenamefont
  {Roberts},\ and\ \citenamefont {Zanotti}}]{Arrington:2006zm}%
  \BibitemOpen
  \bibfield  {author} {\bibinfo {author} {\bibfnamefont {J.}~\bibnamefont
  {Arrington}}, \bibinfo {author} {\bibfnamefont {C.~D.}\ \bibnamefont
  {Roberts}}, \ and\ \bibinfo {author} {\bibfnamefont {J.~M.}\ \bibnamefont
  {Zanotti}},\ }\href {\doibase 10.1088/0954-3899/34/7/S03} {\bibfield
  {journal} {\bibinfo  {journal} {J. Phys. G}\ }\textbf {\bibinfo {volume}
  {34}},\ \bibinfo {pages} {S23} (\bibinfo {year} {2007}{\natexlab{a}})},\
  \Eprint {http://arxiv.org/abs/nucl-th/0611050} {arXiv:nucl-th/0611050}
  \BibitemShut {NoStop}%
\bibitem [{\citenamefont {Perdrisat}\ \emph {et~al.}(2007)\citenamefont
  {Perdrisat}, \citenamefont {Punjabi},\ and\ \citenamefont
  {Vanderhaeghen}}]{Perdrisat:2006hj}%
  \BibitemOpen
  \bibfield  {author} {\bibinfo {author} {\bibfnamefont {C.~F.}\ \bibnamefont
  {Perdrisat}}, \bibinfo {author} {\bibfnamefont {V.}~\bibnamefont {Punjabi}},
  \ and\ \bibinfo {author} {\bibfnamefont {M.}~\bibnamefont {Vanderhaeghen}},\
  }\href {\doibase 10.1016/j.ppnp.2007.05.001} {\bibfield  {journal} {\bibinfo
  {journal} {Prog. Part. Nucl. Phys.}\ }\textbf {\bibinfo {volume} {59}},\
  \bibinfo {pages} {694} (\bibinfo {year} {2007})},\ \Eprint
  {http://arxiv.org/abs/hep-ph/0612014} {arXiv:hep-ph/0612014} \BibitemShut
  {NoStop}%
\bibitem [{\citenamefont {Arrington}\ \emph
  {et~al.}(2011{\natexlab{a}})\citenamefont {Arrington}, \citenamefont
  {de~Jager},\ and\ \citenamefont {Perdrisat}}]{Arrington:2011kb}%
  \BibitemOpen
  \bibfield  {author} {\bibinfo {author} {\bibfnamefont {J.}~\bibnamefont
  {Arrington}}, \bibinfo {author} {\bibfnamefont {K.}~\bibnamefont {de~Jager}},
  \ and\ \bibinfo {author} {\bibfnamefont {C.~F.}\ \bibnamefont {Perdrisat}},\
  }\href {\doibase 10.1088/1742-6596/299/1/012002} {\bibfield  {journal}
  {\bibinfo  {journal} {J. Phys. Conf. Ser.}\ }\textbf {\bibinfo {volume}
  {299}},\ \bibinfo {pages} {012002} (\bibinfo {year} {2011}{\natexlab{a}})},\
  \Eprint {http://arxiv.org/abs/1102.2463} {arXiv:1102.2463 [nucl-ex]}
  \BibitemShut {NoStop}%
\bibitem [{\citenamefont {Jourdan}\ \emph {et~al.}(1997)\citenamefont
  {Jourdan}, \citenamefont {Sick},\ and\ \citenamefont
  {Zhao}}]{Jourdan:1997pw}%
  \BibitemOpen
  \bibfield  {author} {\bibinfo {author} {\bibfnamefont {J.}~\bibnamefont
  {Jourdan}}, \bibinfo {author} {\bibfnamefont {I.}~\bibnamefont {Sick}}, \
  and\ \bibinfo {author} {\bibfnamefont {J.}~\bibnamefont {Zhao}},\ }\href
  {\doibase 10.1103/PhysRevLett.79.5186} {\bibfield  {journal} {\bibinfo
  {journal} {Phys. Rev. Lett.}\ }\textbf {\bibinfo {volume} {79}},\ \bibinfo
  {pages} {5186} (\bibinfo {year} {1997})}\BibitemShut {NoStop}%
\bibitem [{\citenamefont {Bruins}\ \emph {et~al.}(1997)\citenamefont {Bruins}
  \emph {et~al.}}]{Bruins:1997px}%
  \BibitemOpen
  \bibfield  {author} {\bibinfo {author} {\bibfnamefont {E.~E.~W.}\
  \bibnamefont {Bruins}} \emph {et~al.},\ }\href {\doibase
  10.1103/PhysRevLett.79.5187} {\bibfield  {journal} {\bibinfo  {journal}
  {Phys. Rev. Lett.}\ }\textbf {\bibinfo {volume} {79}},\ \bibinfo {pages}
  {5187} (\bibinfo {year} {1997})}\BibitemShut {NoStop}%
\bibitem [{\citenamefont {Arrington}\ \emph {et~al.}(2023)\citenamefont
  {Arrington}, \citenamefont {Cruz-Torres}, \citenamefont {Hague},
  \citenamefont {Kurbany}, \citenamefont {Li}, \citenamefont {Meekins},\ and\
  \citenamefont {Santiesteban}}]{Arrington:2023hht}%
  \BibitemOpen
  \bibfield  {author} {\bibinfo {author} {\bibfnamefont {J.}~\bibnamefont
  {Arrington}}, \bibinfo {author} {\bibfnamefont {R.}~\bibnamefont
  {Cruz-Torres}}, \bibinfo {author} {\bibfnamefont {T.~J.}\ \bibnamefont
  {Hague}}, \bibinfo {author} {\bibfnamefont {L.}~\bibnamefont {Kurbany}},
  \bibinfo {author} {\bibfnamefont {S.}~\bibnamefont {Li}}, \bibinfo {author}
  {\bibfnamefont {D.}~\bibnamefont {Meekins}}, \ and\ \bibinfo {author}
  {\bibfnamefont {N.}~\bibnamefont {Santiesteban}},\ }\href {\doibase
  10.1140/epja/s10050-023-01085-6} {\bibfield  {journal} {\bibinfo  {journal}
  {Eur. Phys. J. A}\ }\textbf {\bibinfo {volume} {59}},\ \bibinfo {pages} {188}
  (\bibinfo {year} {2023})}\BibitemShut {NoStop}%
\bibitem [{\citenamefont {Santiesteban}\ \emph {et~al.}(2021)\citenamefont
  {Santiesteban} \emph {et~al.}}]{Santiesteban:2021lot}%
  \BibitemOpen
  \bibfield  {author} {\bibinfo {author} {\bibfnamefont {S.~N.}\ \bibnamefont
  {Santiesteban}} \emph {et~al.},\ }\href@noop {} {\bibfield  {journal}
  {\bibinfo  {journal} {arXiv:2110.06281}\ } (\bibinfo {year}
  {2021})}\BibitemShut {NoStop}%
\bibitem [{\citenamefont {Alcorn}\ \emph {et~al.}(2004)\citenamefont {Alcorn}
  \emph {et~al.}}]{Alcorn:2004sb}%
  \BibitemOpen
  \bibfield  {author} {\bibinfo {author} {\bibfnamefont {J.}~\bibnamefont
  {Alcorn}} \emph {et~al.},\ }\href {\doibase 10.1016/j.nima.2003.11.415}
  {\bibfield  {journal} {\bibinfo  {journal} {Nucl. Instrum. Meth. A}\ }\textbf
  {\bibinfo {volume} {522}},\ \bibinfo {pages} {294} (\bibinfo {year}
  {2004})}\BibitemShut {NoStop}%
\bibitem [{\citenamefont {Li}\ \emph {et~al.}(2022)\citenamefont {Li} \emph
  {et~al.}}]{Li:2022fhh}%
  \BibitemOpen
  \bibfield  {author} {\bibinfo {author} {\bibfnamefont {S.}~\bibnamefont {Li}}
  \emph {et~al.},\ }\href {\doibase 10.1038/s41586-022-05007-2} {\bibfield
  {journal} {\bibinfo  {journal} {Nature}\ }\textbf {\bibinfo {volume} {609}},\
  \bibinfo {pages} {41} (\bibinfo {year} {2022})}\BibitemShut {NoStop}%
\bibitem [{\citenamefont {Cruz-Torres}\ \emph {et~al.}(2019)\citenamefont
  {Cruz-Torres} \emph {et~al.}}]{cruz-torres19}%
  \BibitemOpen
  \bibfield  {author} {\bibinfo {author} {\bibfnamefont {R.}~\bibnamefont
  {Cruz-Torres}} \emph {et~al.},\ }\href {\doibase
  10.1016/j.physletb.2019.134890} {\bibfield  {journal} {\bibinfo  {journal}
  {Phys. Lett. B}\ }\textbf {\bibinfo {volume} {797}},\ \bibinfo {pages}
  {134890} (\bibinfo {year} {2019})}\BibitemShut {NoStop}%
\bibitem [{\citenamefont {Cruz-Torres}\ \emph {et~al.}(2020)\citenamefont
  {Cruz-Torres} \emph {et~al.}}]{cruz-torres20}%
  \BibitemOpen
  \bibfield  {author} {\bibinfo {author} {\bibfnamefont {R.}~\bibnamefont
  {Cruz-Torres}} \emph {et~al.},\ }\href {\doibase
  10.1103/PhysRevLett.124.212501} {\bibfield  {journal} {\bibinfo  {journal}
  {Phys. Rev. Lett.}\ }\textbf {\bibinfo {volume} {124}},\ \bibinfo {pages}
  {212501} (\bibinfo {year} {2020})}\BibitemShut {NoStop}%
\bibitem [{\citenamefont {Santiesteban}(2021)}]{santiesteban_thesis}%
  \BibitemOpen
  \bibfield  {author} {\bibinfo {author} {\bibfnamefont {S.~N.}\ \bibnamefont
  {Santiesteban}},\ }\href@noop {} {\enquote {\bibinfo {title} {{Ph.D Thesis,
  University New Hampshire}},}\ } (\bibinfo {year} {2021})\BibitemShut
  {NoStop}%
\bibitem [{\citenamefont {Brajuskovic}\ \emph {et~al.}(2013)\citenamefont
  {Brajuskovic}, \citenamefont {O'Connor}, \citenamefont {Holt}, \citenamefont
  {Reneker}, \citenamefont {Meekins},\ and\ \citenamefont
  {Solvignon}}]{target_NIM}%
  \BibitemOpen
  \bibfield  {author} {\bibinfo {author} {\bibfnamefont {B.}~\bibnamefont
  {Brajuskovic}}, \bibinfo {author} {\bibfnamefont {T.}~\bibnamefont
  {O'Connor}}, \bibinfo {author} {\bibfnamefont {R.}~\bibnamefont {Holt}},
  \bibinfo {author} {\bibfnamefont {J.}~\bibnamefont {Reneker}}, \bibinfo
  {author} {\bibfnamefont {D.}~\bibnamefont {Meekins}}, \ and\ \bibinfo
  {author} {\bibfnamefont {P.}~\bibnamefont {Solvignon}},\ }\href {\doibase
  10.1016/j.nima.2013.06.090} {\bibfield  {journal} {\bibinfo  {journal} {Nucl.
  Instrum. Meth. A}\ }\textbf {\bibinfo {volume} {729}},\ \bibinfo {pages}
  {469} (\bibinfo {year} {2013})}\BibitemShut {NoStop}%
\bibitem [{\citenamefont {Santiesteban}\ \emph {et~al.}(2019)\citenamefont
  {Santiesteban} \emph {et~al.}}]{Santiesteban:2018qwi}%
  \BibitemOpen
  \bibfield  {author} {\bibinfo {author} {\bibfnamefont {S.~N.}\ \bibnamefont
  {Santiesteban}} \emph {et~al.},\ }\href {\doibase 10.1016/j.nima.2019.06.025}
  {\bibfield  {journal} {\bibinfo  {journal} {Nucl. Instrum. Meth. A}\ }\textbf
  {\bibinfo {volume} {940}},\ \bibinfo {pages} {351} (\bibinfo {year}
  {2019})},\ \Eprint {http://arxiv.org/abs/1811.12167} {arXiv:1811.12167
  [physics.ins-det]} \BibitemShut {NoStop}%
\bibitem [{\citenamefont {Dasu}\ \emph {et~al.}(1994)\citenamefont {Dasu} \emph
  {et~al.}}]{Dasu:1993vk}%
  \BibitemOpen
  \bibfield  {author} {\bibinfo {author} {\bibfnamefont {S.}~\bibnamefont
  {Dasu}} \emph {et~al.},\ }\href {\doibase 10.1103/PhysRevD.49.5641}
  {\bibfield  {journal} {\bibinfo  {journal} {Phys. Rev. D}\ }\textbf {\bibinfo
  {volume} {49}},\ \bibinfo {pages} {5641} (\bibinfo {year}
  {1994})}\BibitemShut {NoStop}%
\bibitem [{\citenamefont {Arrington}\ \emph {et~al.}(2021)\citenamefont
  {Arrington} \emph {et~al.}}]{Arrington:2021vuu}%
  \BibitemOpen
  \bibfield  {author} {\bibinfo {author} {\bibfnamefont {J.}~\bibnamefont
  {Arrington}} \emph {et~al.},\ }\href@noop {} {\bibfield  {journal} {\bibinfo
  {journal} {Phys. Rev. C}\ }\textbf {\bibinfo {volume} {104}},\ \bibinfo
  {pages} {065203} (\bibinfo {year} {2021})}\BibitemShut {NoStop}%
\bibitem [{\citenamefont {Andreoli}\ \emph {et~al.}(2022)\citenamefont
  {Andreoli}, \citenamefont {Carlson}, \citenamefont {Lovato}, \citenamefont
  {Pastore}, \citenamefont {Rocco},\ and\ \citenamefont
  {Wiringa}}]{Andreoli:2021cxo}%
  \BibitemOpen
  \bibfield  {author} {\bibinfo {author} {\bibfnamefont {L.}~\bibnamefont
  {Andreoli}}, \bibinfo {author} {\bibfnamefont {J.}~\bibnamefont {Carlson}},
  \bibinfo {author} {\bibfnamefont {A.}~\bibnamefont {Lovato}}, \bibinfo
  {author} {\bibfnamefont {S.}~\bibnamefont {Pastore}}, \bibinfo {author}
  {\bibfnamefont {N.}~\bibnamefont {Rocco}}, \ and\ \bibinfo {author}
  {\bibfnamefont {R.~B.}\ \bibnamefont {Wiringa}},\ }\href {\doibase
  10.1103/PhysRevC.105.014002} {\bibfield  {journal} {\bibinfo  {journal}
  {Phys. Rev. C}\ }\textbf {\bibinfo {volume} {105}},\ \bibinfo {pages}
  {014002} (\bibinfo {year} {2022})},\ \Eprint
  {http://arxiv.org/abs/2108.10824} {arXiv:2108.10824 [nucl-th]} \BibitemShut
  {NoStop}%
\bibitem [{\citenamefont {Bosted}\ and\ \citenamefont
  {Mamyan}(2012)}]{Bosted:2012qc}%
  \BibitemOpen
  \bibfield  {author} {\bibinfo {author} {\bibfnamefont {P.~E.}\ \bibnamefont
  {Bosted}}\ and\ \bibinfo {author} {\bibfnamefont {V.}~\bibnamefont
  {Mamyan}},\ }\href@noop {} {\bibfield  {journal} {\bibinfo  {journal}
  {arXiv:1203.2262}\ } (\bibinfo {year} {2012})}\BibitemShut {NoStop}%
\bibitem [{sup()}]{supplemental}%
  \BibitemOpen
  \href@noop {} {}\bibinfo {note} {Supplemental Material}\BibitemShut {NoStop}%
\bibitem [{\citenamefont {Amaro}\ \emph {et~al.}(2022)\citenamefont {Amaro},
  \citenamefont {Martinez-Consentino},\ and\ \citenamefont {Simo}}]{priv_MEC}%
  \BibitemOpen
  \bibfield  {author} {\bibinfo {author} {\bibfnamefont {J.~E.}\ \bibnamefont
  {Amaro}}, \bibinfo {author} {\bibfnamefont {V.~L.}\ \bibnamefont
  {Martinez-Consentino}}, \ and\ \bibinfo {author} {\bibfnamefont {I.~R.}\
  \bibnamefont {Simo}},\ }\href@noop {} {}\bibinfo {howpublished} {private
  communication} (\bibinfo {year} {2022})\BibitemShut {NoStop}%
\bibitem [{\citenamefont {Martinez-Consentino}\ \emph
  {et~al.}(2021)\citenamefont {Martinez-Consentino}, \citenamefont
  {Ruiz~Simo},\ and\ \citenamefont {Amaro}}]{PhysRevC.104.025501}%
  \BibitemOpen
  \bibfield  {author} {\bibinfo {author} {\bibfnamefont {V.~L.}\ \bibnamefont
  {Martinez-Consentino}}, \bibinfo {author} {\bibfnamefont {I.}~\bibnamefont
  {Ruiz~Simo}}, \ and\ \bibinfo {author} {\bibfnamefont {J.~E.}\ \bibnamefont
  {Amaro}},\ }\href {\doibase 10.1103/PhysRevC.104.025501} {\bibfield
  {journal} {\bibinfo  {journal} {Phys. Rev. C}\ }\textbf {\bibinfo {volume}
  {104}},\ \bibinfo {pages} {025501} (\bibinfo {year} {2021})}\BibitemShut
  {NoStop}%
\bibitem [{\citenamefont {Arrington}\ \emph
  {et~al.}(2007{\natexlab{b}})\citenamefont {Arrington}, \citenamefont
  {Melnitchouk},\ and\ \citenamefont {Tjon}}]{arrington:2007ux}%
  \BibitemOpen
  \bibfield  {author} {\bibinfo {author} {\bibfnamefont {J.}~\bibnamefont
  {Arrington}}, \bibinfo {author} {\bibfnamefont {W.}~\bibnamefont
  {Melnitchouk}}, \ and\ \bibinfo {author} {\bibfnamefont {J.~A.}\ \bibnamefont
  {Tjon}},\ }\href {\doibase 10.1103/PhysRevC.76.035205} {\bibfield  {journal}
  {\bibinfo  {journal} {Phys. Rev. C}\ }\textbf {\bibinfo {volume} {76}},\
  \bibinfo {pages} {035205} (\bibinfo {year} {2007}{\natexlab{b}})}\BibitemShut
  {NoStop}%
\bibitem [{\citenamefont {Arrington}\ \emph
  {et~al.}(2011{\natexlab{b}})\citenamefont {Arrington}, \citenamefont
  {Blunden},\ and\ \citenamefont {Melnitchouk}}]{Arrington:2011dn}%
  \BibitemOpen
  \bibfield  {author} {\bibinfo {author} {\bibfnamefont {J.}~\bibnamefont
  {Arrington}}, \bibinfo {author} {\bibfnamefont {P.~G.}\ \bibnamefont
  {Blunden}}, \ and\ \bibinfo {author} {\bibfnamefont {W.}~\bibnamefont
  {Melnitchouk}},\ }\href {\doibase 10.1016/j.ppnp.2011.07.003} {\bibfield
  {journal} {\bibinfo  {journal} {Prog. Part. Nucl. Phys.}\ }\textbf {\bibinfo
  {volume} {66}},\ \bibinfo {pages} {782} (\bibinfo {year}
  {2011}{\natexlab{b}})}\BibitemShut {NoStop}%
\end{thebibliography}%

\end{document}